\documentclass[11pt]{amsart}
\usepackage[margin=1in]{geometry}
\usepackage{amsfonts}
\usepackage{amssymb}
\usepackage[dvips]{graphics}
\usepackage{epsfig}
\usepackage{euscript}
\usepackage{color}

 \theoremstyle{definition}
 
 \theoremstyle{remark}

 \def\idty{{\mathchoice {\mathrm{1\mskip-4mu l}} {\mathrm{1\mskip-4mu l}} %
{\mathrm{1\mskip-4.5mu l}} {\mathrm{1\mskip-5mu l}}}}

\begin{document}

\title[Lieb-Robinson Bounds]{Much Ado About Something\\
Why Lieb-Robinson bounds are useful}

\author[B. Nachtergaele]{Bruno Nachtergaele}
\address{Department of Mathematics\\
University of California, Davis\\
Davis, CA 95616, USA}
\email{bxn@math.ucdavis.edu}

\author[R. Sims]{Robert Sims}
\address{Department of Mathematics\\
University of Arizona\\
Tucson, AZ 85721, USA}
\email{rsims@math.arizona.edu}

\date{Version: \today }
\maketitle

\footnotetext[1]{Copyright \copyright\ 2010 by the authors. This
paper may be reproduced, in its entirety, for non-commercial
purposes.}
\footnotetext[2]{Work supported by National Science Foundation
grants DMS-0757581, DMS-1009502, and DMS-0757424.}
\footnotetext[3]{Article written during a visit of BN to the Institut Mittag-Leffler (Djursholm,
Sweden).}

Understanding many-body dynamics lies at the heart of many fundamental
problems of mathematical physics. Even when one is not directly concerned
with time-dependent phenomena, such as in the study of equilibrium and
non-equilibrium stationary states, one is essentially investigating properties
of the dynamics. It is also possible to approach spectral questions about the
Hamiltonian of a quantum system starting from an analysis of the dynamics it
generates.

The equations of hydrodynamics, the Boltzmann equation, and the Gross-Pitaevskii 
equation, are well-known examples that continue to receive a  lot of attention. In these
examples the aim is to describe the dynamics of a large number, $N$, of identical
particles. More recently, other types of questions have been raised in quantum 
information theory, where one is less interested in the large $N$ limit (although
$N$ may be large) and where one usually does not want to make the assumption
that all degrees of freedom are identical and are subject to identical interactions.
In either case, one is confronted with the task of taming the complexity of many-body
dynamics, and quantum mechanics adds another layer of complexity due to
the role played by entanglement. Fortunately, the interactions in many physical systems 
are either of finite range or their strength decays exponentially (or with a large inverse
power) in the distance between particles. In the past five years Lieb-Robinson bounds
have been shown to be a powerful tool to turn this inherent locality of physical 
systems into useful mathematical estimates. Lieb-Robinson bounds provide an
estimate for the speed of propagation of signals (disturbances) in a spatially
extended system and estimate the magnitude of signals propagating faster
than this speed (the propagation bound is not absolute since we are dealing
with non-relativistic systems). 

Systems defined on a lattice or, more generally, a metric graph, are simpler
than systems of particles in the continuum because the number of degrees
of freedom in a finite region of space can be bounded. So far, Lieb-Robinson 
bounds have only been obtained for such systems but we believe that similar
results are possible for continuum systems and would also be very useful in 
applications. We will only consider lattice systems here, with the term ``lattice''
taken in the broad sense of a discrete set of points. For concreteness and due
to space-limitations, we will restrict ourselves to quantum systems. Similar
considerations have been applied to classical systems such as anharmonic
lattice oscillators \cite{Marchioro78, Butta2007,Raz09} but we will not discuss this further here.
Our basic set-up is then as follows.

Let $\Gamma$ be a set equipped with a metric $d$.
Associated to each $x \in \Gamma$, there is a Hilbert space 
$\mathcal{H}_x$ and a self-adjoint operator $H_x$ defined on a 
dense domain $\mathcal{D}_x \subset \mathcal{H}_x$. 
For example, we could have a harmonic oscillator at each site $x$.
In other applications, consider the isotropic nearest-neighbor 
Heisenberg model in the absence of magnetic fields, we
may have $H_x=0$. The dynamics is then entirely due to 
the interactions, which we will introduce in a moment.
Thus, we have a multi-component quantum system defined 
on $\Gamma$, and for $x,y\in\Gamma$, $d(x,y)$ is interpreted as
the distance between the subsystems located at $x$ and $y$. 
We allow for the possibility that $\Gamma$ is infinite, e.g., 
$\Gamma=\mathbb{Z}^\nu$.
For any finite $\Lambda \subset \Gamma$, the Hilbert space 
of the associated subsystem $\mathcal{H}_{\Lambda}$ and the 
corresponding algebra of observables  $\mathcal{A}_{\Lambda}$ are 
given by
$\mathcal{H}_{\Lambda} = \bigotimes_{x \in \Lambda} \mathcal{H}_x$ and 
$\mathcal{A}_{\Lambda} = \bigotimes \mathcal{B}( \mathcal{H}_x)$,
where $\mathcal{B}( \mathcal{H}_x)$ denotes the bounded linear operators over $\mathcal{H}_x$. 
For an increasing sequence of finite subsets $\Lambda_n\uparrow\Gamma$, the algebra
of local observables is given by the inductive limit 
$\mathcal{A}_{\rm loc}=\cup_{n} \mathcal{A}_{\Lambda_n}$.
This makes sense because $\mathcal{A}_{\Lambda_n}$  is naturally embedded
in $\mathcal{A}_{\Lambda_m}$, for all $m>n$ by identifying $A$ and $A\otimes\idty$,
where $\idty$ is the identity operator on $\mathcal{H}_{\Lambda_m\setminus\Lambda_n}$.
With this identification, for all $A\in\mathcal{A}_{\rm loc}$ we can define the support of
$A$, as the smallest $\Lambda$ such that $A$ belongs to the subalgebra 
$\mathcal{A}_\Lambda$.
The local Hamiltonians for such a system are defined in terms of an interaction. An interaction
is a mapping $\Phi$ from the set of finite subsets of $\Gamma$ into $\mathcal{A}_{\rm loc}$ 
such that for each finite $X \subset \Gamma$, $\Phi(X)^* = \Phi(X) \in \mathcal{A}_X$. Then,
a family of local Hamiltonians $H_{\Lambda}^{\Phi}$,  $\Lambda \subset \Gamma$ finite, 
is defined by
\begin{equation} \label{eq:locham}
H_{\Lambda}^{\Phi} = \sum_{x \in \Lambda} H_x  + \sum_{X \subset \Lambda} \Phi(X) \, .
\end{equation}
When $\Phi$ is understood, we often suppress it in our notation. 
Since the sum in (\ref{eq:locham}) above is finite, and each $\Phi(X)$ is bounded,
$H_{\Lambda}$ is self-adjoint on $\mathcal{H}_{\Lambda}$, and therefore the
Heisenberg dynamics
\begin{equation}
\tau_t^{\Lambda}(A) = e^{itH_{\Lambda}} A e^{-itH_{\Lambda}}  \quad \mbox{for all } 
A \in \mathcal{A}_{\Lambda} \, ,
\end{equation}
is well-defined.

The Lieb-Robinson bounds depend on a combination of properties of $\Gamma$ and $\Phi$.
If $\Gamma$ is infinite, it is necessary to impose a condition, roughly equivalent to 
finite-dimensionality, as follows. We assume that there is a non-increasing, real-valued
function $F: [0, \infty) \to (0, \infty)$, with two properties:
\newline i) {\it uniform integrablility:}
\begin{equation}
\| F \| = \sup_{x \in \Gamma} \sum_{y \in \Gamma} F(d(x,y)) \, < \, \infty \, ,
\end{equation}
\newline ii) {\it convolution property:} there exists a number $C>0$ such that for any pair 
$x,y \in \Gamma$, 
\begin{equation} \label{eq:convc}
\sum_{z \in \Gamma} F(d(x,z)) F(d(z,y) \leq C F(d(x,y)) \, .
\end{equation} 
For the case of $\Gamma = \mathbb{Z}^{\nu}$, one choice of $F$ is given by 
$F(r) = (1+r)^{ \nu + 1}$. Then the convolution property holds with 
$C = 2^{ \nu +1} \sum_{x \in \Gamma}F(|x|)$. Note that  one can
assume $C=1$ without loss of generality (replace $F$ by $C^{-1}F$.)
An important observation is that if there exists a function $F$ on $\Gamma$
satisfying i) and ii), then for any $\mu \geq 0$, the function
$F_{\mu}$ defined by setting $F_{\mu}(r) = e^{-\mu r}F(r)$ also
satisfies i) and ii) with $\| F_{\mu} \| \leq \| F \|$ and $C_{\mu} \leq C$.
For any $\mu \geq 0$,  we denote by $\mathcal{B}_{\mu}( \Gamma)$ the set of interactions 
$\Phi$ for which
\begin{equation}
\| \Phi \|_{\mu} = \sup_{x,y \in \Gamma} \frac{1}{F_{\mu}(d(x,y))} \sum_{\stackrel{X \subset \Gamma:}{x,y \in X}} \| \Phi(X) \| \, < \infty \, .
\end{equation}

If $\Phi \in \mathcal{B}_{\mu}( \Gamma)$, then a Lieb-Robinson bound of the form
\begin{equation}
\left\| \left[ \tau_t^{\Lambda}(A), B \right] \right\| \leq  2   \| A \| \| B \| C_\mu^{-1} \, 
 (e^{2 C_{\mu} \| \Phi \|_{\mu} |t|} -1) \sum_{x \in X}\sum_{y \in Y} F_{\mu}(d(x,y))
\end{equation}
holds for all $A \in \mathcal{A}_X$, $B \in \mathcal{A}_Y$, $X\cap Y=\emptyset$,
and $t \in \mathbb{R}$. If $\mu>0$, the double sum can be bounded by an exponentially
decaying factor of the form $C\Vert F\Vert e^{- \mu d(X,Y)}$, which leads
to a version of the bound in the familiar form:
\begin{equation}
\left\| \left[ \tau_t^{\Lambda}(A), B \right] \right\| \leq 2  \| A \| \| B \| C
e^{-\mu( d(X,Y) -v |t|)} \, .
\label{LRoriginal}\end{equation}
Here, $v=2\mu^{-1} C_{\mu} \| \Phi \|_{\mu}$ is the Lieb-Robinson velocity 
and one can take $C= C_{\mu}^{-1} \min(\vert X\vert, \vert Y \vert)$ or in the case of interactions of 
finite range $R$, $C=R C_{\mu}^{-1} \min(\vert \partial X\vert, \vert \partial Y \vert)$, where 
$\vert \partial Z\vert$ denotes the size of the boundary of $Z$.
A brief discussion of the literature is in order here. The original bound of the form
(\ref{LRoriginal}), albeit in a slightly more restricted setup, appeared in 1972 in \cite{lieb1972}
(see also \cite{bratteli1997}). In the following three decades, apart from a few works 
considering classical lattice oscillators \cite{Marchioro78} and a calculations for specific 
models \cite{radin1978}, applications and extensions of Lieb-Robinson bounds received 
little attention.
This changed in 2004 with Hastings' work \cite{hastings2004a} on the multi-dimensional
Lieb-Schultz-Mattis Theorem \cite{lieb1961}. In his paper,  Hastings used Lieb-Robinson 
bounds in combination with his ``quasi-adiabatic continuation'' technique to analyze 
properties of ground states of quantum lattice systems and their excitations.
For a self-contained presentation of quasi-adiabatic continuation see \cite{lppl}. This
technique and Lieb-Robinson bounds were subsequently extended and used in new 
applications by Hastings and collaborators \cite{hastings05,hastings06,bravyi:2006,%
hastings2007,hastings09,hastings2010,bravyi:2010a,bravyi:2010b} and other authors 
\cite{NachS06,Nach06,Nach07,Burrell07,Osborne07,cramer:2008,Erdos08,NachS09,%
Nach09,Hamza09,PS09,Raz09,Nach10,NachS10,PS10,Poulin10}.
 
The new applications motivated extensions of the Lieb-Robinson bounds to
more general systems as well as improvements on the basic estimates. We will mention some of these
newer results below but space limitations will not allow us to explicitly mention
all the results that have appeared in the past few years. First, however, we clarify the relation
between bounds on commutators and the support of observables.

Let $\mathcal{H}_1$ and $\mathcal{H}_2$ be Hilbert spaces and suppose there is an $A\in\mathcal{B}(
\mathcal{H}_1\otimes\mathcal{H}_2)$ such that for all $B\in\mathcal{B}(\mathcal{H}_2)$ we have
$\Vert [A,\idty\otimes B]\Vert \leq \epsilon \Vert B\Vert$. Then, there exists $A_\epsilon
\in\mathcal{B}(\mathcal{H}_1)$ such that 
$\Vert A - A_\epsilon\Vert \leq \epsilon$. In fact, if $\mathcal{H}_2$ is finite-dimensional $A_\epsilon$ 
can be obtained using the partial trace of $A$ over $\mathcal{H}_2$, but the result also holds 
for bounded observables on infinite-dimensional Hilbert spaces \cite{lppl}. For fixed $A \in \mathcal{A}_X$,
one can apply this result  with $\tau^\Lambda_t(A)$ taking the role of $A$. By letting $\mathcal{H}_1=\mathcal{H}_{X_r}$, 
where $X_r = \{y\in\Lambda, d(X,y)\leq v\vert t\vert + r\}$ and 
$\mathcal{H}_2 = \mathcal{H}_{\Lambda\setminus X_r}$, we see that (\ref{LRoriginal}) gives 
a bound for the error one makes if one replaces $\tau^\Lambda_t(A)$ by 
its best approximation with  support contained in $X_r$. Therefore, it is not surprising
that Lieb-Robinson bounds can be used to prove the existence of the dynamics in the 
thermodynamic limit \cite{bratteli1997}. While Lieb-Robinson bounds do not always
give the best possible estimates, in some cases they do, and moreover, they have been used to prove 
the only known results for anharmonic lattices \cite{Nach10,Amour09}.

Another fundamental result that directly relies on the locality properties expressed by 
Lieb-Robinson bounds is the Exponential Clustering Theorem.
It states that Hamiltonians with a non-vanishing spectral gap have ground states with 
exponentially decaying spatial correlations. Although the corresponding result in relativistic
quantum field theory has been known for a long time, the non-relativistic version, with
the Lieb-Robinson velocity playing the role of the speed of light, was only proved
in 2006 \cite{NachS06,hastings06}. The correlation length $\xi$ satisfies 
$\xi\leq 2v/\gamma +1/\mu$, where $v$ is the Lieb-Robinson velocity, $\gamma$
is the spectral gap above the ground state, and $\mu$ is the parameter measuring 
the exponential rate of decay of the interaction.
When one uses a more recent version of the 
Lieb-Robinson bounds \cite{NachS09}, the exponential clustering for gapped
ground states $\omega$ of one-dimensional systems with short-range interactions,
can be shown to hold in the following strong form: 
$$
\vert \omega(AB)-\omega(A)\omega(B)\vert\leq c\Vert A\Vert\, \Vert B\Vert \exp{(-d(X,Y)/\xi)},
$$
where $X$ and $Y$ are the supports of
$A$ and $B$, respectively. Note that only the distance between the supports and not their 
size appears in the estimate, a feature that was exploited by Matsui in his investigation
of the split property for quantum spin chains \cite{matsui:2010}. 

The Area Law, the conjecture that the entropy of the restrictions of 
gapped ground states of quantum lattice models to a finite volume $\Lambda$
grows no faster than a quantity proportional to the surface area of $\Lambda$, has
been proved by Hastings for one-dimensional systems in \cite{hastings2007}.
In his paper Lieb-Robinson bounds are used to derive an approximate factorization 
property of the density matrices of gapped ground states. Such a result can be
generalized to higher dimensions \cite{Hamza09}, but it does not, by itself, suffice to 
prove the Area Law in this context. This issue remains a topic of active 
investigation (see, e.g., \cite{debeaudrap10}).

The estimate of the correlation length in terms of the gap plays an important role in several 
other applications of Lieb-Robinson bounds, including the multi-dimensional 
Lieb-Schultz-Mattis (LSM) theorem. A precise statement of the general multi-dimensional LSM 
theorem would be too long to fit in the space alotted, but since it was one of the first non-trivial
applications of Lieb-Robinson bounds, it deserves to be discussed here. For concreteness
consider the spin-$S$ nearest-neighbor isotropic quantum Heisenberg antiferromagnet with 
$S=1/2,3/2,5/2,\ldots$, defined on a finite subset of $\mathbb{Z}^\nu$ of the form
$[1,2L]\times [2L+1]^{\nu-1}$, and with periodic boundary conditions in the first coordinate. 
It is known that this model has a unique ground state \cite{lieb1962}. The LSM theorem
\cite{hastings2004a,Nach07} then provides a bound for the energy of the first excited
state:
$$
E_1-E_0 \leq C\frac{\log L}{L}\, ,
$$
where $C$ is a constant of order 1, only depending on the dimension and the coupling
constant. The proof is by showing that a gap {\em larger} than the bound claimed above
would allow one to construct variational states of {\em lower} energy, which is a contradiction.
The properties of the variational states, the estimate of their energy and the proof of
orthogonality to the ground state, rely on Lieb-Robinson bounds in an essential way.
See \cite[Section 5]{NachS09} for a more detailed outline of the complete proof.
Here we just mention that Lieb-Robinson bounds allow one to show that, as long as the 
spectral gap is not too small, local perturbations added to a Hamiltonian modify the ground 
state only in a neighborhood of the perturbation. See \cite{lppl} for a general proof
of this property.

In an impressive application of the adiabatic continuation technique and Lieb-Robinson
bounds Hastings and Michalakis proved the quantization of the Hall conductance
for a general class of models of interacting fermions on a lattice in \cite{hastings09}.
An extension to the fractional quantum Hall effect is also discussed in this work.

Bravyi, Hastings, and Michalakis \cite{bravyi:2010a,bravyi:2010b} have shown that 
topological order in the ground states of a class of Hamiltonians 
that are the sum of commuting short-range terms, such as Kitaev's toric code model
\cite{kitaev2003}, is stable under arbitrary sufficiently small short-range perturbations.
This can be regarded as another instance where Lieb-Robinson bounds are used
to show that local perturbations have only local effects, and therefore cannot destroy
a global property such as topological order.

In this short review we have only considered Hamiltonian quantum dynamics. 
We just note that Lieb-Robinson bounds have also been derived for irreversible 
dynamics described by semigroups with a generator of Lindblad form \cite{Poulin10}.

The complexity of quantum dynamics and its tendency to create entangled states
are a barrier to our intuitive understanding of many of the most interesting physical 
phenomena. Good mathematical results that elucidate the structure behind this
complexity are essential to aid our understanding.  We hope to have convinced the
reader that Lieb-Robinson bounds are a good example of this kind of mathematical 
result.

\end{document}